# Improvements of Polya Upper Bound for Cumulative Standard Normal Distribution and Related Functions

Omar M. Eidous, Professor of Statistics
omarm@yu.edu.jo

Department of Statistics – Yarmouk University
Irbid – Jordan

**Abstract**

Although there is an extensive literature on the upper bound for cumulative standard normal distribution $\Phi(x)$, there are relatively not sharp for all values of the interested argument $x$. The aim of this paper is to establish a sharp upper bound for $\Phi(x)$, in the sense that its maximum absolute difference from $\Phi(x)$ is less than $5.785 \times 10^{-5}$ for all values of $x \geq 0$. The established bound improves the well-known Polya upper bound and it can be used as an approximation for $\Phi(x)$ itself with a very satisfactory accuracy. Numerical comparisons between the proposed upper bound and some other existing upper bounds have been achieved, which show that the proposed bound is tighter than alternative bounds found in the literature.

**Keywords:** Bonds for CDF of standard normal; Polya bound; Error function; Q-function; Maximum absolute error.

**Mathematics Subject Classifications (2000) 33B20 · 26D15**

## 1. Introduction

The cumulative standard normal (Gaussian) distribution $\Phi(x)$ and its related functions are appeared frequently in applied mathematics, statistics and in analytical studies of communication systems over fading channels (Simon, 2006). The cumulative distribution function (CDF) of a standard normal random variable $X$ is defined as,

$$\Phi(x) = \int_{-\infty}^{x} \phi_X(t)dt$$

where $\phi_X(t)$ is the standard normal probability density function of a random variable $X$, which is given by,

$$\phi_X(t) = \frac{\exp(-t^2/2)}{\sqrt{2\pi}}, \quad -\infty < t < \infty.$$

It is very difficult to handle $\Phi(x)$ mathematically because it cannot be expressed in closed form. Therefore, several works have been developed to approximate and find more compact lower and upper bounds for $\Phi(x)$ and its related special functions like the Q-function and error function.

The probability that a standard normal random variable $X$ exceeds a positive value $x$ is known as Q-function and is defined by,

$$Q(x) = \int_{x}^{\infty} \phi_X(t)dt,$$

which is related to $\Phi(x)$ by,

$$Q(x) = 1 - \Phi(x).$$

The other two well-known special functions that can be described in term of $\Phi(x)$ and $Q(x)$ are the error function $erf(x)$ and the complementary error function $ercf(x)$, which are given by,

$$\begin{aligned} erf(x) &= \frac{2}{\sqrt{\pi}} \int_{0}^{x} \exp(-t^2)\, dt \\ &= 2\Phi(\sqrt{2}x) - 1 \\ &= 1 - 2Q(\sqrt{2}x) \end{aligned}$$

and

$$\begin{aligned} erfc(x) &= 1 - erf(x) \\ &= 2 - 2\Phi(\sqrt{2}x) \\ &= 2Q(\sqrt{2}x). \end{aligned}$$

The Q-function and error function are widely used in Bit Error Rate (BER) analysis of communication systems (see Trigui *et al.*, 2021 and the references theirin). Due to the relationship between $\Phi(x)$ and the above functions along with the Mills' ratio (Fan, 2013), it makes sense that the results would find application in the broader fields such as statistical computations, applied statistics, mathematical models in biology, mathematical physics and diffusion theory (see Bercu, 2020 and Lipoth *et al.*, 2022). Eidous and Abu-Shareefa (2020) have reviewed 45 approximations for $\Phi(x)$ reported in literature from 1945 to 2019. They also provided nine accurate approximations for $\Phi(x)$. All of these approximations can be modified simply to approximate the functions $Q(x)$, $erf(x)$ and $ercf(x)$.

In this paper we will focus our attention on the function $\Phi(x)$. Instead of approximating $\Phi(x)$, this paper proposes a new upper bound for $\Phi(x)$, which improves the well-known Polya upper bound and the proposed bound can be used as an approximation for $\Phi(x)$ with very small absolute error not exceed $5.785 \times 10^{-5}$. The proposed upper bound is compared with some alternative upper bounds found in the literature, which demonstrates the accuracy of the proposed one for all argument values greater than zero. Finally, all results related to $\Phi(x)$ can be generalized and used –with some modification- for the other functions $Q(x)$, $erf(x)$ and $ercf(x)$.

## 2. Upper Bounds for $\Phi(x)$

The integral in $\Phi(x)$ -as well as the integrals in $Q(x)$, $erf(x)$ and $ercf(x)$- cannot be evaluated analytically, so values for these functions are computed via approximations and often available in tables. Therefore, some bounds and approximations for these functions have been developed and suggested in the literature as given below.

- Polya (1949) considered the following simple form of upper bound for $\Phi(x)$ (for $x > 0$),

$$\Phi(x) \leq \Phi_{PO}(x)$$

where

$$\Phi_{PO}(x) = \frac{1}{2}\sqrt{1 - \exp\left(-\frac{2x^2}{\pi}\right)}$$

is the Polya upper bound. Iacono (2021) pointed out that $\Phi_{PO}(x)$ is fairly accurate.

- Kouba (2006) derived the following upper bound for $\Phi(x)$,

$$\Phi(x) \leq \Phi_{KO}(x)$$

where $\Phi_{KO}(x) = 1 - \frac{\phi(x)}{t}$ and $t = \sqrt{1 + \left(\frac{x}{2}\right)^2} + \frac{x}{2}$.

- The resemblance between $\Phi(x)$ and $\tanh(x)$ has inspired a more recent result by Alzer (2010), whose upper bound for $\Phi(x)$ is given as follows (for $x > 0$),

$$\Phi(x) \leq \Phi_{AL}(x)$$

where,

$$\Phi_{AL}(x) = 0.5 + \frac{1.0407\, Tanh\left(\sqrt{\frac{2}{\pi}}\, x\right)}{2}.$$

- Abreu (2012) gave the following upper bound for $\Phi(x)$ (for $x > 0$),

$$\Phi(x) \leq \Phi_{AB}(x)$$

where,

$$\Phi_{AB}(x) = 1 - \frac{\exp(-x^2)}{12} - \frac{\phi(x)}{1+x}.$$

- Neumann (2013) gave the following upper bound for $\Phi(x)$ (for $x > 0$),

$$\Phi(x) \leq \Phi_{NE}(x)$$

where,

$$\Phi_{NE}(x) = \frac{1}{2} + \frac{x}{3} \frac{2 + \exp(-x^2/2)}{\sqrt{2\pi}}.$$

- Yang *et. al.* (2018) improved Neumann bound $\Phi_{NE}(x)$ by suggested the following bound, for $x > 0$,

$$\Phi(x) \leq \Phi_{YA}(x)$$

where,

$$\Phi_{YA}(x) = \frac{1}{2} + \frac{x}{9} \frac{4 + 5\exp(-3x^2/10)}{\sqrt{2\pi}}.$$

- Let $y = x/\sqrt{2}$ then Bercu (2020) gave the following upper bound for $\Phi(x)$ when $0 \leq y \leq 4.418$

$$\Phi(x) \leq \Phi_{BE}(x)$$

where

$$\Phi_{BE}(x) = \frac{1}{2} + \frac{1}{\sqrt{\pi}} \frac{113400\, y}{29y^8 - 660y^6 + 1260y^4 + 37800y^2 + 113400}.$$

With regard to the above upper bounds of $\Phi(x)$, a limitation shared by all of these works is that they do not exhibit a tight upper bound for all values of argument $x > 0$. For instance and as the numerical results in Section (4) shown, $\Phi_{KO}(x)$ is very tightness for large values of $x$ but it is less sharp for small values of $x$. The converse is true for the upper bound $\Phi_{YA}(x)$. In light of this, an important contribution of this work is to provide a new tightness and sharp upper bound of $\Phi(x)$ for all values of $x \geq 0$. In particular, the new suggested bound improved the Polya upper bound, which has been a topic of substantial interest in contemporary mathematics and statistics.

### 3. Proposed Upper Bound of $\Phi(x)$

The following lemma gives the proposed upper bound for $\Phi(x)$.

**Lemma 1:** Let

$$\Phi_{EI}(x) = \frac{1}{2}\sqrt{1 - \exp\left(-\frac{2x^2}{\pi}\left(1 + \frac{(-\pi + 3)x^2}{3\pi} + \left(\frac{7}{90} + \frac{40001}{30000\pi^2} - \frac{2}{3\pi}\right)x^4\right)\right)}$$

then, for $x \geq 0$, $\Phi_{EI}(x)$ is an upper bound for $\Phi(x)$. That is,

$$\Phi(x) \leq \Phi_{EI}(x).$$

**Proof:** To verify that $\Phi(x) \leq \Phi_{EI}(x)$, it is enough to show that $\Phi_{EI}(x) - \Phi(x) \geq 0$. The derivative of $h(x) = \Phi_{EI}(x) - \Phi(x)$ is,

$$h'(x) = -\frac{e^{-\frac{x^2}{2}}}{\sqrt{2\pi}} + \frac{x(120003x^4 - 60000\pi x^2(-1 + x^2) + 1000\pi^2(30 - 20x^2 + 7x^4))\exp(r(x))}{30000\sqrt{1 - \pi^3 \exp(r(x))}}$$

where

$$r(x) = \frac{1}{\pi}\left(-\frac{40001x^6}{15000\pi^2} - 2x^2 + \frac{30x^4 - 7x^6}{45} + \frac{-6x^4 + 4x^6}{3\pi}\right).$$

By using Mathematica, Ver 11, it is found that there is only one root for $h'(x) = 0$, which is $x \approx 2.86991$. This can be illustrated in Graph (1), which gives the plot of $h'(x)$.

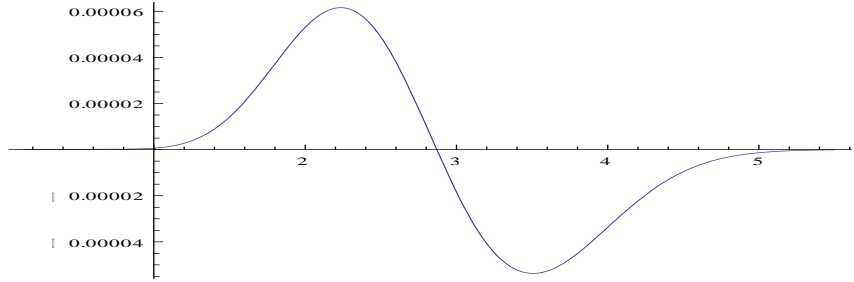

**Graph (1)**. Plot of $h'(x)$.

Also, the plot of $h(x)$ is given in Graph (2) below. From these two graphs and by recognize the one root of $h'(x) = 0$, it is clear that $h(x)$ is increasing on $x \in [0, 2.86991]$ and decreasing on $x \in [2.86991, \infty)$ with maximum value of $h(x) = 5.784 \times 10^{-5}$ occurs at $x = 2.86991$, i.e. $h(2.86991) \approx 5.784 \times 10^{-5}$.

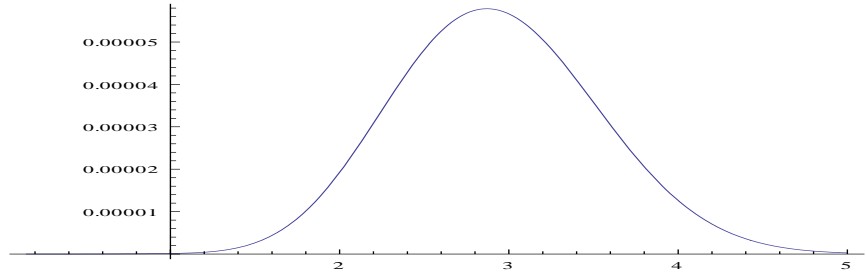

**Graph (2)**. Plot of $h(x) = \Phi_{EI}(x) - \Phi(x)$.

On one hand and since $h(x)$ is increasing on $x \in [0, 2.86991]$ then its smallest possible value throughout this interval occurs at $x = 0$, which is clearly 0, i.e. $h(0) = 0$. On the other hand and because $h(x)$ decreases on the interval $[2.86991, \infty)$ then it is smallest possible value occurs as $x \to \infty$. Since the coefficient of $x^4$ in the exponent term of $u(x)$ is $\frac{7}{90} + \frac{40001}{30000\pi^2} - \frac{2}{3\pi} \approx 0.000666 > 0$, then

$$\lim_{x \to \infty} u(x) = \lim_{x \to \infty} \frac{1}{2}\sqrt{1 - \exp\left(-\frac{2x^2}{\pi}\left(1 + \frac{(-\pi + 3)x^2}{3\pi} + \left(\frac{7}{90} + \frac{40001}{30000\pi^2} - \frac{2}{3\pi}\right)x^4\right)\right)} = \frac{1}{2},$$

and hence,

$$\lim_{x \to \infty}(\Phi_{EI}(x) - \Phi(x)) = 0.$$

This completes the proof.

**Notes:**
1. Since the Q-function, $Q(x) = 1 - \Phi(x)$ then a proposed lower bound for $Q(x)$ based on $\Phi_{EI}(x)$ is given by,

$$Q(x) \geq 1 - \Phi_{EI}(x).$$

Also, a proposed upper bound for the error function $erf(x)$ based on $\Phi_{EI}(x)$ is,
$$erf(x) \leq 2\Phi_{EI}(\sqrt{2}x) - 1.$$

2. Since $\frac{7}{90} + \frac{40001}{30000\pi^2} - \frac{2}{3\pi} \approx 0.000666$ and $\frac{(-\pi+3)}{3\pi} \approx -0.015023$, then the proposed inequality can be expressed more simply as follows,

$$\Phi(x) \leq \frac{1}{2}\sqrt{1 - \exp\left(-\frac{2x^2}{\pi}(1 - 0.015023\, x^2 + 0.000666\, x^4)\right)}.$$

3. The proposed upper bound $\Phi_{EI}(x)$ of $\Phi(x)$ can be used as an approximation for $\Phi(x)$, i.e.
$$\Phi(x) \cong \Phi_{EI}(x), \quad x \geq 0$$
with maximum absolute error equals $5.784 \times 10^{-5}$, i.e.
$$\max_{x \geq 0}|\Phi_{EI}(x) - \Phi(x)| \cong 5.784 \times 10^{-5}.$$

In addition and as an approximation for $\Phi(x)$, further improvements can be introduced for the above formula of $\Phi_{EI}(x)$ by using the following approximation,
$$\Phi(x) \cong \Phi_{EI}^*(x),$$
where,
$$\Phi_{EI}^*(x) = \frac{1}{2}\sqrt{1 - \exp\left(-\frac{2x^2}{\pi}(1 - 0.01506\, x^2 + 0.00063\, x^4)\right)}.$$

In this case, the maximum absolute error between $\Phi_{EI}^*(x)$ and $\Phi(x)$ is $3.1677 \times 10^{-5}$ which is less than that of $\Phi_{EI}(x)$. In fact, the maximum absolute error between $\Phi_{EI}(x)$ and $\Phi(x)$ equals (approximately) 1.826 of the maximum absolute error between $\Phi_{EI}^*(x)$ and $\Phi(x)$. That is,
$$\max_{x \geq 0}|\Phi_{EI}(x) - \Phi(x)| \cong 1.826 \max_{x \geq 0}|\Phi_{EI}^*(x) - \Phi(x)|.$$

It is worthwhile to mention here that $\Phi_{EI}^*(x)$ is neither an upper nor lower bound for $\Phi(x)$. Graph (3) shows the plot of $h^*(x) = \Phi_{EI}^*(x) - \Phi(x)$.

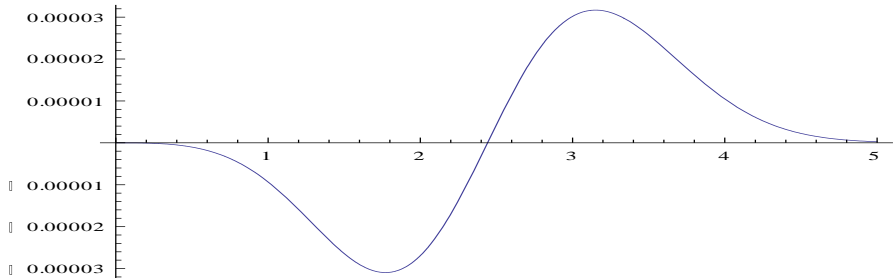

**Graph (3)**. Plot of $h^*(x) = \Phi_{EI}^*(x) - \Phi(x)$.

4. Based on Note (2), the Q-function and error function can be approximated by,
$$Q(x) \cong 1 - \Phi_{EI}(x) \quad (\text{or } Q(x) \cong 1 - \Phi_{EI}^*(x))$$
and
$$erf(x) \cong 2\Phi_{EI}(\sqrt{2}x) - 1 \quad (\text{or } erf(x) \cong 2\Phi_{EI}^*(\sqrt{2}x) - 1).$$

**Lemma 2:** For $0 \leq x \leq \sqrt{\frac{\pi-3}{\frac{7\pi}{30}+\frac{40001}{10000\pi}-2}} \cong 4.74915$,

$$\Phi_{EI}(x) \leq \Phi_{PO}(x).$$

Lemma 2 states that, for $0 \leq x \leq 4.74915$, the proposed upper bound $\Phi_{EI}(x)$ is more closer to $\Phi(x)$ than Polya upper bound $\Phi_{PO}(x)$.

For $x > 4.74915$, it is found that the maximum error of $\Phi_{EI}(x)$ is less than $9.16 \times 10^{-7}$, while it is less than $9.15 \times 10^{-7}$ for the bound $\Phi_{PO}(x)$ (see the results of Table 1).

**Proof of Lemma 2:** It is simple to show that (for $x \geq 0$)

$$\Phi_{EI}(x) \leq \Phi_{PO}(x)$$

If and only if,

$$\frac{1}{2}\sqrt{1 - \exp\left(-\frac{2x^2}{\pi}\left(1 + \frac{(-\pi+3)x^2}{3\pi} + \left(\frac{7}{90} + \frac{40001}{30000\pi^2} - \frac{2}{3\pi}\right)x^4\right)\right)} \leq \frac{1}{2}\sqrt{1 - \exp\left(-\frac{2x^2}{\pi}\right)}$$

If and only if,

$$\exp\left(-\frac{2x^2}{\pi}\left(1 + \frac{(-\pi+3)x^2}{3\pi} + \left(\frac{7}{90} + \frac{40001}{30000\pi^2} - \frac{2}{3\pi}\right)x^4\right)\right) \geq \exp\left(-\frac{2x^2}{\pi}\right)$$

If and only if,

$$1 + \frac{(-\pi+3)x^2}{3\pi} + \left(\frac{7}{90} + \frac{40001}{30000\pi^2} - \frac{2}{3\pi}\right)x^4 \leq 1$$

If and only if,

$$\frac{-\pi+3}{3\pi} + \left(\frac{7}{90} + \frac{40001}{30000\pi^2} - \frac{2}{3\pi}\right)x^2 \leq 0$$

If and only if,

$$\left(\frac{7}{90} + \frac{40001}{30000\pi^2} - \frac{2}{3\pi}\right)x^2 \leq \frac{\pi-3}{3\pi}$$

If and only if,

$$\left(\frac{7}{90} + \frac{40001}{30000\pi^2} - \frac{2}{3\pi}\right)x^2 \leq \frac{\pi-3}{3\pi}$$

If and only if,

$$x \leq \sqrt{\frac{\pi-3}{\frac{7\pi}{30}+\frac{40001}{10000\pi}-2}} \cong 4.74915.$$

This completes the proof.

### 5. Numerical Comparisons

Let $h_U(x)$ be the error function between the upper bounds of $\Phi(x)$ and the exact $\Phi(x)$. That is, $h_U(x) = \Phi_U(x) - \Phi(x)$, where $U$ stands for $KO$ (Kouba, 2006), $AL$ (Alzer, 2010), $AB$ (Abreu, 2012), $NE$ (Neumann, 2013), $YA$ (Yang et al., 2018), $BE$ (Bercu, 2020) and $PO$ (Polya, 1949) (see Section 2). The error function for the proposed upper bound is $h_{EI}(x) = \Phi_{EI}(x) - \Phi(x)$ (see Section 3). The entries of Table (1) represent the function $h(x)$ for only some values of $x$ varies from 0.1 to 8.5, in spite of the fact that all bounds considered in this paper (except the bound $\Phi_{BE}(x)$) can be indeed applied to any $x \geq 0$. The focus of our analysis is on small and moderate argument values since our proposed bound conjunction with Polya bound are very precise for large argument values (greater than 8.0, say). By examining the numerical results of Table (1), the following can be concluded:

1. The proposed upper bound is very accurate for small values of $x$ ($x < 1.9, say$) and for large values of $x$ ($x > 3.9, say$). It still accurate for the other values of $x$ comparing with the other upper bounds considered in this study.
2. One can anticipate that our proposal in conjunction with, for instance, $\Phi_{YA}(x)$ and $\Phi_{BE}(x)$ being highly accurate for small argument values. However, the tightness of $\Phi_{YA}(x)$ and $\Phi_{BE}(x)$ are not as good as the proposed bound for moderate to large argument values. It is worthwhile to mention here that the bound $\Phi_{BE}(x)$ is developed for $0 \leq y = x/\sqrt{2} \leq 4.418$ ($i.e.$ $0 \leq x \leq 6.248$) as stated in Section (2). The numerical study depicted this case, which is clear from the negative values of $h_{BE}(x)$ for $x \geq 6.5$. The disadvantage of using the bound $\Phi_{BE}(x)$ for only a sub-interval of $[0, \infty)$ reduces the possibility of its use in various fields of analysis.
3. As the proposed bound $\Phi_{KO}(x)$, the two bounds $\Phi_{KO}(x)$ and $\Phi_{AB}(x)$ are very tightness to the exact $\Phi(x)$ as $x$ increases beyond $x = 3$. Moreover, one can conclude that each bound becomes even more accurate as $x \rightarrow \infty$. However, the numerical values show clearly the significance improvements of the proposed bound over $\Phi_{KO}(x)$ and $\Phi_{AB}(x)$ for $0 < x < 2.5$.
4. By taking into account all the considered values of the argument $x$ in this numerical study, one can conclude that the tightness of the two bounds $\Phi_{AL}(x)$ and $\Phi_{NE}(x)$ is not as good as the other bounds.
5. By comparing the proposed upper bound, $\Phi_{EI}(x)$ with Polya upper bound, $\Phi_{PO}(x)$, the numerical results of Table (1) show clearly that $\Phi_{EI}(x)$ is more tighter than $\Phi_{PO}(x)$ for $x < 5$. For $x \geq 5$, the two bounds are very closed to the exact $\Phi(x)$ with maximum error less than $2.6 \times 10^{-7}$ for $\Phi_{PO}(x)$ and less than $2.8 \times 10^{-7}$ for $\Phi_{EI}(x)$.

**6. Conclusions**

In this paper, we have proposed a new upper bound $\Phi_{EI}(x)$ for the $\Phi(x)$. The proposed $\Phi_{EI}(x)$ has simple expression and it is more tightness to exact $\Phi(x)$ than most of the other bounds considered in this study for small to moderate arguments. In addition, its accuracy for large argument is still very high despite that some existing bounds are more accurate than it. This justifies the possibility of use the proposed bound as an approximation for $\Phi(x)$ as discussed in Section (3) of this paper. It is worthwhile to mention here that all results related $\Phi_{EI}(x)$ can be simply modified and then used for the well-known special functions Q-function and error function. Finally, the simplicity and accuracy of the proposed bound indicate the wider and high possibility of its application in different fields of analysis involving $\Phi(x)$ or any related functions Q-function, error function and complementary error function.

**Table (1).** The values of error function $h_U(x) = \Phi_U(x) - \Phi(x)$ for some values of $x$.

| $x$ | $h_{KO}(x)$ | $h_{AL}(x)$ | $h_{AB}(x)$ | $h_{NE}(x)$ | $h_{YA}(x)$ | $h_{BE}(x)$ | $h_{PO}(x)$ | $h_{EI}(x)$ |
|---|---|---|---|---|---|---|---|---|
| 0.1 | 1.11×10⁻² | 1.60×10⁻³ | 1.68×10⁻² | 6.63×10⁻⁸ | 1.90×10⁻¹¹ | 0.00 | 2.98×10⁻⁶ | 4.63×10⁻¹¹ |
| 0.3 | 9.13×10⁻³ | 4.32×10⁻³ | 1.26×10⁻² | 1.58×10⁻⁵ | 4.08×10⁻⁸ | 3.11×10⁻¹² | 7.72×10⁻⁵ | 1.48×10⁻⁹ |
| 0.5 | 6.77×10⁻³ | 5.77×10⁻³ | 8.93×10⁻³ | 1.96×10⁻⁴ | 1.41×10⁻⁶ | 8.05×10⁻¹⁰ | 3.29×10⁻⁴ | 1.29×10⁻⁸ |
| 0.7 | 4.66×10⁻³ | 5.72×10⁻³ | 7.23×10⁻³ | 9.96×10⁻⁴ | 1.41×10⁻⁵ | 2.96×10⁻⁸ | 7.96×10⁻⁴ | 4.21×10⁻⁸ |
| 0.9 | 3.03×10⁻³ | 4.46×10⁻³ | 6.94×10⁻³ | 3.25×10⁻³ | 7.66×10⁻⁵ | 4.16×10⁻⁷ | 1.43×10⁻³ | 8.92×10⁻⁸ |
| 1.1 | 1.88×10⁻³ | 2.64×10⁻³ | 7.08×10⁻³ | 8.10×10⁻³ | 2.87×10⁻⁴ | 3.25×10⁻⁶ | 2.11×10⁻³ | 2.29×10⁻⁷ |
| 1.3 | 1.12×10⁻³ | 1.01×10⁻³ | 6.92×10⁻³ | 1.68×10⁻² | 8.38×10⁻⁴ | 1.71×10⁻⁵ | 2.70×10⁻³ | 8.14×10⁻⁷ |
| 1.5 | 6.40×10⁻⁴ | 9.57×10⁻⁵ | 6.22×10⁻³ | 3.05×10⁻² | 2.04×10⁻³ | 6.80×10⁻⁵ | 3.06×10⁻³ | 2.65×10⁻⁶ |
| 1.7 | 3.53×10⁻⁴ | 1.63×10⁻⁴ | 5.10×10⁻³ | 5.00×10⁻² | 4.31×10⁻³ | 2.17×10⁻⁴ | 3.14×10⁻³ | 6.85×10⁻⁶ |
| 1.9 | 1.88×10⁻⁴ | 1.19×10⁻³ | 3.84×10⁻³ | 7.56×10⁻² | 8.18×10⁻³ | 5.86×10⁻⁴ | 2.94×10⁻³ | 1.43×10⁻⁵ |
| 2.1 | 9.67×10⁻⁵ | 2.98×10⁻³ | 2.66×10⁻³ | 1.07×10⁻¹ | 1.42×10⁻² | 1.38×10⁻³ | 2.54×10⁻³ | 2.48×10⁻⁵ |
| 2.3 | 4.82×10⁻⁵ | 5.23×10⁻³ | 1.72×10⁻³ | 1.44×10⁻¹ | 2.28×10⁻² | 2.92×10⁻³ | 2.03×10⁻³ | 3.70×10⁻⁵ |
| 2.5 | 2.32×10⁻⁵ | 7.65×10⁻³ | 1.04×10⁻³ | 1.86×10⁻¹ | 3.45×10⁻² | 5.65×10⁻³ | 1.51×10⁻³ | 4.82×10⁻⁵ |
| 2.7 | 1.08×10⁻⁵ | 1.00×10⁻² | 5.94×10⁻⁴ | 2.31×10⁻¹ | 4.94×10⁻² | 1.01×10⁻² | 1.05×10⁻³ | 5.57×10⁻⁵ |
| 2.9 | 4.88×10⁻⁶ | 1.21×10⁻² | 3.21×10⁻⁴ | 2.79×10⁻¹ | 6.76×10⁻² | 1.71×10⁻² | 6.82×10⁻⁴ | 5.78×10⁻⁵ |
| 3.1 | 2.13×10⁻⁶ | 1.4×10⁻² | 1.65×10⁻⁴ | 3.29×10⁻¹ | 8.91×10⁻² | 2.75×10⁻² | 4.17×10⁻⁴ | 5.41×10⁻⁵ |
| 3.3 | 9.00×10⁻⁷ | 1.55×10⁻² | 8.13×10⁻⁵ | 3.80×10⁻¹ | 1.13×10⁻¹ | 4.23×10⁻² | 2.40×10⁻⁴ | 4.61×10⁻⁵ |
| 3.5 | 3.68×10⁻⁷ | 1.67×10⁻² | 3.83×10⁻⁵ | 4.32×10⁻¹ | 1.4×10⁻¹ | 6.30×10⁻² | 1.30×10⁻⁴ | 3.58×10⁻⁵ |
| 3.7 | 1.45×10⁻⁷ | 1.76×10⁻² | 1.73×10⁻⁵ | 4.85×10⁻¹ | 1.7×10⁻¹ | 9.12×10⁻² | 6.68×10⁻⁵ | 2.53×10⁻⁵ |
| 3.9 | 5.56×10⁻⁸ | 1.83×10⁻² | 7.53×10⁻⁶ | 5.38×10⁻¹ | 2.01×10⁻¹ | 1.29×10⁻¹ | 3.25×10⁻⁵ | 1.63×10⁻⁵ |
| 4.1 | 2.05×10⁻⁸ | 1.89×10⁻² | 3.15×10⁻⁶ | 5.91×10⁻¹ | 2.33×10⁻¹ | 1.80×10⁻¹ | 1.50×10⁻⁵ | 9.57×10⁻⁶ |
| 4.4 | 4.33×10⁻⁹ | 1.94×10⁻² | 7.93×10⁻⁷ | 6.70×10⁻¹ | 2.83×10⁻¹ | 2.86×10⁻¹ | 4.30×10⁻⁶ | 3.60×10⁻⁶ |
| 4.7 | 8.43×10⁻¹⁰ | 1.98×10⁻² | 1.83×10⁻⁷ | 7.50×10⁻¹ | 3.35×10⁻¹ | 4.37×10⁻¹ | 1.11×10⁻⁶ | 1.09×10⁻⁶ |
| 5.0 | 1.52×10⁻¹⁰ | 2.00×10⁻² | 3.89×10⁻⁸ | 8.30×10⁻¹ | 3.87×10⁻¹ | 6.21×10⁻¹ | 2.56×10⁻⁷ | 2.71×10⁻⁷ |
| 5.5 | 7.33×10⁻¹² | 2.02×10⁻² | 2.42×10⁻⁹ | 9.63×10⁻¹ | 4.75×10⁻¹ | 7.00×10⁻¹ | 1.79×10⁻⁸ | 1.89×10⁻⁸ |
| 6.0 | 2.83×10⁻¹³ | 2.03×10⁻² | 1.19×10⁻¹⁰ | 1.10 | 5.64×10⁻¹ | 1.83×10⁻¹ | 9.59×10⁻¹⁰ | 9.87×10⁻¹⁰ |
| 6.5 | 8.66×10⁻¹⁵ | 2.03×10⁻² | 4.57×10⁻¹² | 1.23 | 6.53×10⁻¹ | -2.14×10⁻¹ | 3.96×10⁻¹¹ | 4.02×10⁻¹¹ |
| 7.0 | 2.22×10⁻¹⁶ | 2.03×10⁻² | 1.38×10⁻¹³ | 1.36 | 7.41×10⁻¹ | -3.75×10⁻¹ | 1.27×10⁻¹² | 1.28×10⁻¹² |
| 7.5 | 1.11×10⁻¹⁶ | 2.03×10⁻² | 3.33×10⁻¹⁵ | 1.49 | 8.3×10⁻¹ | -4.40×10⁻¹ | 3.19×10⁻¹⁴ | 3.19×10⁻¹⁴ |
| 8.0 | 0.00 | 2.03×10⁻² | 1.11×10⁻¹⁶ | 1.63 | 9.18×10⁻¹ | -4.68×10⁻¹ | 6.66×10⁻¹⁶ | 6.66×10⁻¹⁶ |
| 8.5 | 0.00 | 2.04×10⁻² | 0.00 | 1.76 | 1.01 | -4.82×10⁻¹ | 0.00 | 0.00 |